\documentclass{elsart}

\usepackage{graphicx}
\usepackage{epsfig,array}
\usepackage{booktabs}

\newcommand{\beq}{\begin{equation}}
\newcommand{\eeq}{\end{equation}}
\newcommand{\beqa}{\begin{eqnarray}}
\newcommand{\eeqa}{\end{eqnarray}}

\newcommand{\sla}[1]%
        {\kern .25em\raise.18ex\hbox{$/$}\kern-.75em #1}
\newcommand{\mybar}[1]%
        {\kern 0.8pt\overline{\kern -0.8pt#1\kern -0.8pt}\kern 0.8pt}

\begin{document} 
\begin{frontmatter}
\title{Higgs-Mediated $e\rightarrow \mu$ transitions
in II Higgs doublet Model and Supersymmetry}
\author[romeII]{P.Paradisi}
\address[romeII]{University of Rome ``Tor Vergata'' and INFN sez. RomaII, 
	Via della Ricerca Scientifica 1, 
	I-00133 Rome}

\begin{abstract}

We study the phenomenology of the $e-\mu$ lepton flavor violation (LFV) 
in a general two Higgs Doublet Model (2HDM) including the supersymmetric case.
We compute the decay rate expressions of $\mu\rightarrow e\gamma$, 
$\mu\rightarrow eee$, and $\mu\rightarrow e$ conversion in nuclei at 
two loop level. In particular, it is shown that $\mu\rightarrow e\gamma$ 
is generally the most sensitive channel to probe Higgs-mediated LFV.
The correlations among the decay rates of the above processes
are also discussed.

\end{abstract}

\end{frontmatter}

\section{Introduction}
\label{sec:introduction}

The Supersymmetric (SUSY) extension of the Standard Model (SM) is one of
the most promising candidate for physics beyond the SM.
Besides direct searches for SUSY particles, it is also important
to analyze implications of such a theory in the low-energy phenomena,
through virtual effects of SUSY particles.
Lepton flavor violating (LFV) processes are excellent candidates to explore
such virtual SUSY effects from low energy experiments \cite{roberts}.
In fact, in a SM framework with massive neutrinos, all the LFV transitions are
expected at a very suppressed level, very far from any future and reasonable
experimental resolutions.
On the other hand, the observation of neutrino oscillation have established
the existence of lepton family number violation.\\
This mixing is expected to be manifested also in the charged lepton sector
through the observation of rare decay processes such as
$\mu\rightarrow e\gamma$, $\tau\rightarrow \mu\gamma$ etc.
Any experimental signals of such a process would clearly indicate the presence
of a non standard mechanism.
In a supersymmetric (SUSY) framework, new direct sources of flavor violation appear, provided the presence of off-diagonal soft terms in the
slepton mass matrices and in the trilinear couplings \cite{fbam}.
In practice, flavor violation would originate from any misalignment 
between fermion and sfermion mass eigenstates.
LFV processes arise at one loop level through the exchange of
neutralinos (charginos) and charged sleptons (sneutrinos).
The amount of the LFV is regulated by a Super-GIM mechanism that can be
much less severe than in the non supersymmetric case 
\cite{Masierorew,textures,masivemp}
\footnote{As recently shown in ref.\cite{mfvl}, some of these effects are common
to many extensions of the SM, even to non-susy scenarios,
and can be described in a general way in terms of an effective field theory.
Moreover, in the context of general $SU(2)_L\times U(1)_Y$ seesaw scenarios,
large LFV effects can be induced by the exchange of left-handed and/or
right-handed neutral singlets \cite{pillfv}.}.\\
Another potential source of LFV in models such as the minimal
supersymmetric standard model (MSSM) is the Higgs sector.\\
In fact, extensions of the Standard Model containing more than one Higgs 
doublet generally allow flavor-violating couplings of the neutral Higgs bosons
with fermions. Such couplings, if unsuppressed, will lead to large
flavor-changing neutral currents in direct opposition to experiments.\\
The MSSM avoids these dangerous couplings at the tree level segregating
the quark and Higgs fields so that one Higgs $(H_u)$ can couple only to
up-type quarks while the other $(H_d)$ couples only to d-type.
Within unbroken supersymmetry this division is completely natural,
in fact, it is required by the holomorphy of the superpotential.
However, after supersymmetry is broken, couplings of the form $QU_cH_d$
and $QD_cH_u$ are generated at one loop \cite{hrs}.\\
As shown in ref.\cite{posptgb,bkq}, the presence of these loop-induced 
non holomorphic couplings also leads to the appearance of flavor-changing 
couplings of the neutral Higgs bosons that are particularly relevant at 
large values of $\tan\beta$.
As a natural consequence, a variety of flavor-changing processes such as 
$B^0 \rightarrow\mu^{+}\mu^{-}$ \cite{bkq}, $\bar{B^0}-B^0$ 
\cite{isiburapila}, $K\rightarrow \pi\nu\bar{\nu}$ \cite{kpnntgb} etc. 
is generated 
\footnote{For a recent and detailed analysis of the $B$ physics phenomenology
at large $\tan\beta$ within the Minimal Flavor Violating framework, see
\cite{hintstgb}.}.
Higgs-mediated FCNC can have sizable effects also in the lepton sector \cite{bkl}: 
given a source of nonholomorphic couplings,
and LFV among the sleptons, Higgs-mediated LFV is unavoidable.\\
These effects have been widely discussed in the recent literature 
both in a generic 2HDM \cite{chang1,chang,hmio} and in supersymmetry 
\cite{hmio,sher,brignole,ellis,kitano} frameworks.
In particular, it has been shown that a tree level Higgs-exchange leads to 
$\tau\rightarrow l_jl_kl_k$ \cite{bkl}, $\tau\rightarrow l_j\eta$ \cite{sher},
$B^{0}\rightarrow l_j\tau$ \cite{ellis}
and $\mu\rightarrow e$ conversion in Nuclei \cite{kitano}.\\
Recently, it was pointed out that Higgs-mediated LFV effects can also 
generate violations of lepton universality at the $1\%$ level in the
$R=\Gamma(K\rightarrow e\nu)/\Gamma(K\rightarrow \mu\nu)$ ratio \cite{kl2}.\\
Moreover, Higgs-mediated FCNC can have a sizable impact also in loop-induced 
processes, such as $\tau\rightarrow l_j\gamma$ \cite{hmio}.\\
In this letter, we investigate the effects of Higgs mediated LFV in the $e-\mu$
transitions both in a generic two Higgs Doublet Model (2HDM) and in Supersymmetry.
We evaluate analytical expressions and correlations for the rates of 
$\mu\rightarrow e\gamma$, $\mu\rightarrow eee$ and
$\mu\rightarrow e$ conversion in nuclei at two loop level,
establishing which the most promising channels to detect LFV signals are.

\section{LFV in the Higgs Sector}

SM extensions containing more than one Higgs doublet generally allow 
flavor-violating couplings of the neutral Higgs bosons with fermions.
Such couplings, if unsuppressed, will lead to large flavor-changing
neutral currents in direct opposition to experiments.
The possible solution to this problem involves an assumption about the
Yukawa structure of the model. A discrete symmetry can be invoked to allow a given fermion type to couple to a single Higgs doublet, and in such case FCNC's are absent at tree level. In particular, when a single Higgs field gives masses to both types of fermions the resulting model is referred as 2HDM-I. 
On the other hand, when each type of fermion couples to a different 
Higgs doublet the model is said 2HDM-II.\\
In the following, we will assume a scenario where the type-II 2HDM structure 
is not protected by any symmetry and is broken by loop effects 
(this occurs, for instance, in the MSSM).\\
Let us consider the Yukawa interactions for charged leptons,
including the radiatively induced LFV terms:

\beq
\mathcal{-L}\simeq
\overline{l}_{Ri}Y_{l_{i}}H_1\overline{L_i}+
\overline{l}_{Ri}\left(Y_{l_{i}}\Delta_{L}^{ij}+Y_{l_{j}}\Delta_{R}^{ij}\right)H_2
\overline{L_j} + h.c.
\eeq
where $H_1$ and $H_2$ are the scalar doublets, $l_{Ri}$ are lepton singlet for right handed fermions, $L_{k}$ denote the lepton doublets and $Y_{l_{k}}$ are 
the Yukawa couplings.

In the mass-eigenstate basis for both leptons and Higgs bosons,
the effective flavor-violating interactions are described by the 
four dimension operators:
\beqa
\mathcal{-L}&\simeq&(2G_{F}^2)^{\frac{1}{4}}\frac{m_{l_i}} {c^2_{\beta}}
\left(\Delta^{ij}_{L}\overline{l}^i_R l^j_L+\Delta^{ij}_{R}\overline{l}^i_L l^j_R \right)
\left(c_{\beta-\alpha}h^0-s_{\beta-\alpha}H^0-iA^0 \right)\nonumber\\\nonumber\\
&+&
(8G_{F}^2)^{\frac{1}{4}}\frac{m_{l_i}} {c^2_{\beta}}
\left(\Delta^{ij}_{L}\overline{l}^i_R \nu^j_L+\Delta^{ij}_{R}\nu^i_L\overline{l}^j_R\right)
H^{\pm} + h.c.
\eeqa
where $\alpha$ is the mixing angle between the CP-even Higgs bosons $h_0$ and
$H_0$, $A_0$ is the physical CP-odd boson, $H^{\pm}$ are the physical
charged Higgs-bosons and $t_{\beta}$ is the ratio of the vacuum expectation value for the two Higgs (where we adopt the notation, $c_{x},s_{x}\!=\!\cos x,\sin x$ and
$t_{x}\!=\!\tan x$).
Irrespective to the mechanism of the high energy theories generating the LFV,
we treat the $\Delta^{ij}_{L,R}$ terms in a model independent way.
In order to constrain the $\Delta^{ij}_{L,R}$ parameters,
we impose that their contributions to LFV processes
do not exceed the experimental bounds \cite{hmio}.\\
On the other hand, there are several models with a specific ansatz about the flavor-changing couplings.
For instance, the famous multi-Higgs-doublet models proposed by Cheng and Sher \cite{chengsher} predict that the LFV couplings of all the neutral Higgs bosons 
with the fermions have the form $Hf_if_j \sim \sqrt{m_im_j}$.

In Supersymmetry, the $\Delta^{ij}$ terms are induced at one loop level 
by the exchange of gauginos and sleptons, provided a source of slepton mixing.
In the so called MI approximation, the expressions of $\Delta^{ij}_{L,R}$ are given by
\beqa
\label{deltal}
 \Delta^{ij}_{L} =
&-&\frac{\alpha_{1}}{4\pi}\mu M_1 \delta^{ij}_{LL} m_{L}^2
\left[
I^{'} (M_1^2, m_{R}^2, m_{L}^2)+\frac{1}{2} I^{'} (M_1^2, \mu^2, m_{L}^2)
\right]+\nonumber\\\nonumber\\
&+& \frac{3}{2} \frac{\alpha_{2}}{4\pi} \mu M_2 \delta^{ij}_{LL} m_{L}^2
I^{'} (M_2^2, \mu^2, m_{L}^2)\ ,
\eeqa
\beq
\label{deltar}
\Delta^{ij}_{R}=
\frac{\alpha_{1}}{4\pi}\mu M_1 m^{2}_{R} \delta^{ij}_{RR}
\left[I^{'}\!(M^{2}_{1},\mu^2,m^{2}_{R})\!-\!(\mu\!\leftrightarrow\! m_{L})
\right]
\eeq
respectively, where $\mu$ is the the Higgs mixing parameter, $M_{1,2}$ 
are the gaugino masses and $m^{2}_{L(R)}$ stands for the left-left 
(right-right) slepton mass matrix entry. The LFV mass insertions (MIs), i.e.
$\delta^{3\ell}_{XX}\!=\!({\tilde m}^2_{\ell})^{3\ell}_{XX}/m^{2}_{X}$ $(X=L,R)$,
are the off-diagonal flavor changing entries of the slepton mass matrix.
The loop function $I^{'}(x,y,z)$ is such that $I^{'}(x,y,z)= dI(x,y,z)/d z$, 
where $I(x,y,z)$ refers to the standard three point one-loop integral which 
has mass dimension -2
\begin{eqnarray}
 I_3 (x, y, z)=
\frac{xy \log (x/y) + yz \log (y/z) + zx \log (z/x)}
{(x-y)(z-y)(z-x)}\,.
\end{eqnarray}
The above expressions, i.e. the Eqs.\ref{deltal},\ref{deltar}, depend only on 
the ratio of the susy mass scales and they do not decouple for large $m_{SUSY}$.
As first shown in Ref.\cite{brignole}, both $\Delta^{ij}_{R}$ and
$\Delta^{ij}_{L}$ couplings suffer from strong cancellations in certain 
regions of the parameter space due to destructive interferences 
among various contributions.
For instance, from Eq.\ref{deltar} it is clear that,
in the $\Delta^{ij}_{R}$ case, such cancellations happen if $\mu=m_L$.

In the SUSY seesaw model, the MIs of the slepton mass matrix appear 
in the left-handed sleptons through the neutrino Yukawa interactions.
The superpotential of the lepton sector is given by
$W = Y_e H_1 l_{R}^{c} L+Y_\nu H_2 N^c L+(1/2) M_N N^c N^c$,
where $N^c$ is the superfields corresponding to the right-handed neutrinos.
The neutrino mass matrix is obtained by integrating out the heavy
right-handed neutrinos as $ m_{\nu} = (Y_\nu^T
M_N^{-1} Y_\nu ) v^2 \sin^2 \beta/2$, where $v$ is the vacuum
expectation value (VEV) of the Higgs field ($v = 246$ GeV).
The correct size of the neutrino masses is obtained for
right-handed neutrinos as heavy as $10^{14}$ GeV
for $f_\nu \sim O(1)$.
The Yukawa coupling $Y_\nu$ violates the lepton flavor conservation
and this violation is communicated to the slepton mass matrix at low-energy.
The renormalization group equation (RGE) running effect induces the 
following off-diagonal components in the left-handed slepton mass matrix
\beq
(\tilde{m}_{\ell_{L}}^2)_{ij}
\simeq 
- \frac{1}{8 \pi^2}m_0^2 (3 + a^2) 
\bigg(Y^{\dagger}_\nu\log\frac{M_{\rm GUT}}{M_N}Y_\nu\bigg)_{ij}
\ ,
\eeq
where the SUSY breaking parameters $m_0$ and $a$ stand for
the scalar mass and the trilinear scalar coupling
at the GUT scale, respectively.

Given our ignorance about the mixings in $Y_\nu$,
we consider two extremal benchmark cases as discussed, within
the $SO(10)$ framework, in \cite{masivemp}.
As a minimal mixing case we take the one in which the neutrino and the
up-quark Yukawa unify at the high scale, so that the mixing is given by the
CKM matrix; this case is named `CKM--case'. 
As a maximal mixing scenario we take the one in which the observed 
neutrino mixing is coming entirely from the neutrino Yukawa matrix, so that 
$Y_\nu = U_{PMNS} \cdot Y^{\mathrm{diag}}_u$, where $U_{PMNS}$ is the
Pontecorvo--Maki--Nakagawa--Sakata matrix; in this case the 
unknown $U_{e3}$ PMNS matrix element turns out to be crucial in
evaluating the size of LFV effects. The maximal case is named
`PMNS--case'. As regards the $\delta_{LL}^{21}$ MI one obtains 
\cite{masivemp}
\beqa
\label{ckmcase}
\delta_{LL}^{21}&=& -{3  \over 8 \pi^2}~
Y_t^2 V_{td} V_{ts} \ln{M_{X} \over M_{R_3}}\,\,\qquad\qquad \rm{CKM-case} 
\\
\label{mnscase}
\delta_{LL}^{21}&=& -{3  \over 8 \pi^2}~
Y_t^2 U_{e3} U_{\mu3} \ln{M_X \over M_{R_3}}\qquad\qquad \rm{PMNS-case}\,.
\eeqa
where we set $a=0$. So, in the CKM--case, it turns out that 
$\delta_{LL}^{21}\simeq 3\cdot 10^{-5}$
while in the PMNS--case, taking $U_{e3}=0.07$ at about half of the current 
CHOOZ bound, we get $\delta_{LL}^{21}\simeq 10^{-2}$.

\section{$\bf{e-\mu}$ transitions in the non-decoupling limit}

In this section, we will analyze $e-\mu$ transitions through the study of
$\mu\rightarrow e\gamma$, $\mu\rightarrow eee$ and $\mu\rightarrow e$
conversion in nuclei in the non-decoupling limit of a 2HDM, where
$s_{\beta-\alpha}\!=\!0$ and $t_{\beta}$ is large.
In particular, we derive analytical expressions and correlations
for the examined branching ratios in order to establish which
the most promising channels to detect Higgs mediated LFV are.
$\mu\rightarrow e\gamma$ is generated by a dipole operator arising,
at least, from one loop Higgs exchange.
However, higgs mediated dipole transitions imply three chirality
flips: two in the Yukawa vertices and one in the lepton propagator.
This strong suppression can be overcome at higher order level.
Going to two loop level, one has to pay the typical $g^2/16\pi^2$ price
but one can replace light fermion masses from yukawa vertices with
heavy fermion (boson) masses circulating in the second loop 
\cite{chang1,barrzee}.
In this case, the virtual higgs boson couples only once to the lepton
line, inducing the needed chirality flip.
As a result, two loop amplitudes provide the major effects and
we find that $Br(\mu\rightarrow e\gamma)$ is given by
\beqa
\label{main2h}
Br(\mu\rightarrow e \gamma)\simeq
\frac{3}{8}\frac{\alpha_{el}}{\pi}\frac{m^4_{\mu}}{M^4_{h,A}}
\Delta_{21}^{2}t_{\beta}^6
&\bigg[&\pm\log{\frac{m^2_{\mu}}{M^2_{h,A}}}
-\frac{2\alpha_{el}}{\pi}\bigg(\frac{m^{2}_W}{m^{2}_{\mu}}\bigg)
\frac{F(a_{W})}{t_{\beta}}+
\nonumber\\\nonumber\\
&\pm&
\frac{\alpha_{el}}{\pi}\sum_{f= b, \tau}\!\!\!N_fq^2_{f}
\bigg(\frac{m^2_{f}}{m^2_{\mu}}\bigg)
\bigg(\log{\frac{m^2_{f}}{M^2_{h,A}}}\bigg)^{2}
\bigg]^2
\eeqa
where $N_{\tau, b}= 1,3$, $q_f$ is the electric charge of the fermion $f$ and
$a_{W}=m^{2}_{W}/m^{2}_{h}$. The term proportional to $F(a_W)$ arises
from two loop effects induced by Barr-Zee type diagrams \cite{barrzee} 
with a $W$ boson exchange. The loop function $F(z)$ is given by
\beq
F(z)\simeq 3f(z)+\frac{23}{4}g(z)+\frac{f(z)-g(z)}{2z}
\eeq
with the Barr-Zee loop integrals given by:
\beq
g(z) = \frac{1}{4}\int_0^1 dx \frac{\log{(z/x(1-x))}}{z-x(1-x)}\,,
\eeq
\beq
f(z) = \frac{1}{4}\int_0^1 dx \frac{1-2x(1-x)\log{(z/x(1-x))}}{z-x(1-x)}.
\eeq
For $z\ll 1$ it turns out that:
\beq
\label{fapprox}
F(z)\sim \frac{35}{16}(\log z)^{2}+\frac{\log z+2}{4z}.
\eeq
The first term of Eq.\ref{main2h} refers to one loop contributions
while the last term arises from two loop effects induced by fermionic loops.
In the computation, we retained only the $h^0\gamma\gamma$ effective vertex
neglecting the $(1-4\sin^2_W)$ suppressed contributions arising from the
$h^0Z\gamma$ vertex.\\
To get a feeling on the relative size among different contributions,
we note that two loop fermionic (bosonic) amplitudes are enhanced by an
$m^{2}_{f}/m^{2}_{\mu}$ ($m^{2}_{W}\cot\beta/m^{2}_{\mu}$) factor
with respect to the one loop amplitude.
In fact, one gets heavy fermionic (bosonic) masses both from the fermionic (bosonic) propagators and from the $H\bar{f} f \sim m_f t_{\beta}$
($H W W \sim m_W$) couplings.
Two loop effects generated by the top quark are generally subdominant.
In fact, bearing in mind that any Model II 2HDM predicts
that $H\bar{t} t \sim m_t/t_{\beta}$ and noting that
the top amplitude isn't enhanced by large logarithm factors
one finds naively that
$$
\frac{A^{b}}{A^{t}}\sim\frac{q^{2}_{b}}{q^{2}_{t}}
\frac{m^{2}_{b}t^{2}_{\beta}}{m^{2}_{top}}
\left(\log\frac{m^{2}_{b}}{m^{2}_{h}}\right)^{\!2},
$$
where $A_{t,b}$ stands for the top and bottom two loop amplitudes.
Since the Higgs mediated LFV is relevant only at large $t_{\beta}\geq 30$,
it is clear that $\tau$ and $b$ contributions are dominant.\\
Moreover, from Eqs.\ref{main2h}-\ref{fapprox} it is straightforward 
to check that two loop effects are largely dominated by the $W$ exchange 
instead of the exchange of heavy fermions.
A possible exception arises only if $m_{A}\ll m_{h}$.
In fact, bearing in mind that pseudoscalar bosons do not couple to a $W$ pair,
it turns out that two loop $W$ effects are sensitive only to scalar mediation,
in contrast to the fermionic case.
At this point, we proceed to consider the contributions to
$\mu\rightarrow eee$ and $\mu Al\rightarrow e Al$.
We find that $\mu\rightarrow eee$ is completely dominated by the photonic
$\mu\rightarrow e\gamma^{*}$ dipole amplitude so that
$Br(\mu\rightarrow eee)\simeq \alpha_{em} Br(\mu\rightarrow e\gamma)$.
On the other hand, $\mu\rightarrow e $ conversion in Nuclei gets the major effects by
the scalar operator through the tree level Higgs exchange that leads to the following expression for $Br(\mu Al\rightarrow e Al)$:
\beqa
Br(\mu Al\rightarrow e Al) &\simeq& 
\,1.8\times10^{-4}\,
\frac{m^{7}_{\mu}m^{2}_{p}}{v^{4}m^{4}_{h}\omega^{Al}_{capt}}\,
\Delta^{2}_{21}t_{\beta}^{6}\,,
\eeqa
where $\omega^{Al}_{capt}\simeq 0.7054\cdot 10^{6}sec^{-1}$.
In fact, in contrast to $\mu \to 3e$, that is suppressed by the electron mass through the $H(A)\bar{e}e \sim m_e$ coupling, $\mu N\rightarrow e N$ is not suppressed by the light constituent quark $m_u$ and $m_d$ but only by the nucleon masses, because the Higgs-boson coupling to the nucleon is shown to be characterized by the nucleon mass using the conformal anomaly relation \cite{shifman}. In particular, the most important contribution turns out to come from the exchange of the scalar Higgs boson $h$ and $H$ which couples to the strange quark \cite{corsetti}
\footnote{As discussed in \cite{kitano}, the coherent $\mu-e$ conversion process,
where the initial and final nuclei are in the ground state,
is expected to be enhanced by a factor of $O(Z)$ (where $Z$ is the atomic number)
compared to incoherent transition processes.
Since the initial and final states are the same,
the elements $\langle N | \bar{p} p | N \rangle$
and $\langle N | \bar{n} n | N \rangle$ are nothing but
the proton and the neutron densities in a nucleus in the non-relativistic
limit of nucleons. In this limit, the other matrix elements
$\langle N | \bar{p} \gamma_5 p | N \rangle$ and
$\langle N | \bar{n} \gamma_5 n | N \rangle$ vanish.
Therefore, in the coherent $\mu-e$ conversion process,
the dominant contributions come from the exchange of
$h$ and $H$, not $A$.}.
Moreover, from a previous analysis \cite{hmio},
we know that $\mu\rightarrow e\gamma^{*}$ (chirality conserving)
monopole amplitudes are generally subdominant compared to
(chirality flipping) dipole effects.
In addition, the enhancement mechanism induced by Barr-Zee type
diagrams is effective only for chirality flipping operators so, in the following,
we will disregard chirality conserving one loop effects.
Let us derive now the approximate relations among
$\mu Al\rightarrow e Al$, $\mu\rightarrow e\gamma$ and $\mu\rightarrow eee$
branching ratios
\beqa
\frac{Br(\mu\rightarrow e \gamma)}{Br(\mu Al\rightarrow e Al)} &\simeq& 
10^{\,2}\,\left(\frac{F(a_W)}{\tan\beta}\right)^{2}\,\,,\,\,\,\,
\frac{Br(\mu\rightarrow eee)}{Br(\mu \rightarrow e \gamma)}\simeq\alpha_{el}
\eeqa
where the approximate expression for $F(a_W)$ is given by Eq.\ref{fapprox}.
In the above equations we retained only dominant two loop effects
arising from $W$ exchange.
The exact behavior for the examined processes is reported in fig.1
where we can see that $\mu\rightarrow e\gamma$ gets the largest branching ratio
except for a region around $m_H\sim 700 \rm{Gev}$ where strong cancellations
among two loop effects sink its size.
For a detailed discussion about the origin of these cancellations
and their connection with non-decoupling properties of two loop
$W$ amplitude, see ref.\cite{chang1}.
On the other hand, $\mu\rightarrow e\gamma$ amplitude can receive large one loop contributions by a double LFV source, namely
by $(\Delta^{21})_{eff.}\!=\!\Delta^{23}\Delta^{31}$ and therefore,
the resulting $Br(\mu\rightarrow e\gamma)$ is:
\beqa
\label{2delta}
Br(\mu\rightarrow e \gamma)
&\simeq&
\frac{3}{8}\frac{\alpha_{el}}{\pi}
\bigg(\frac{m^2_{\tau}}{M^2_{h,A}}\bigg)^{2}t_{\beta}^{8}
\bigg[
\bigg(\pm\log{\frac{m^2_{\tau}}{M^2_{h,A}}}\!+\!\frac{4(-5)}{3}\bigg)
\Delta_{L}^{23}\Delta_{L}^{31}+\nonumber\\
&\pm&
\bigg(\frac{m_{\tau}}{m_{\mu}}\bigg)
\bigg(\log{\frac{m^2_{\tau}}{M^2_{h,A}}}\!+\!\frac{3}{2}\bigg)
\Delta_{R}^{23}\Delta_{L}^{31}
\bigg]^{2}+(L\!\leftrightarrow\!R).
\eeqa
Now, in contrast to one loop contributions with a single LFV coupling
(see the first term of Eq.\ref{main2h}), it is always possible to pick 
up $m_{\tau}$ instead of $m_{\mu}$ both at the LFV Yukawa vertices and 
at the fermion propagator.
However, if the LFV couplings are generated radiatively
(as it happens for instance in a Susy framework),
the above enhancement is modulated by the loop suppression.
In practice, the dominance of one loop effects (with two LFV couplings)
over two loop effects (with one LFV coupling) depends on the specific model
we are treating, namely on the size of $\Delta_{ij}$ terms.
Assuming that the contributions with a double source of LFV 
(see Eq.\ref{2delta}) dominate over those with a single LFV source 
(see Eq.\ref{main2h}), the following ratios are expected:
\beq
\label{corr2delta}
\frac{Br(\mu Al\rightarrow e Al)}{Br(\mu \rightarrow e \gamma)}\simeq
\frac{Br(\mu\rightarrow eee)}
{Br(\mu \rightarrow e \gamma)}\simeq\alpha_{el}.
\eeq
The $\mu\rightarrow e\gamma^{*}$ dominance in $Br(\mu Al\rightarrow e Al)$
and $Br(\mu\rightarrow eee)$ is the reason of the above correlations.
On the other hand, the same correlations are expected, for instance,
in a Susy framework with gaugino mediated LFV and then,
the predictions of Eq.\ref{corr2delta} prevent us from distinguishing
between the two scenarios.
Possible deviations from Eq.\ref{corr2delta} can arise only through
tree level Higgs exchange effects to $\mu Al\rightarrow e Al$.

\section{$\bf{e-\mu}$ transitions in the decoupling limit}

The decoupling limit of a 2HDM is a particularly appealing scenario in that
it is achieved by Supersymmetry. In this context, the higgs bosons masses are
nearly degenerate $m_{A}\simeq m_{H}\simeq m_{H^{\pm}}$
being the mass splitting of order $\mathcal{O}(m^2_Z/m_{A})$ and,
in addition, it turns out that $c_{\beta-\alpha}\!=\!0$ and 
$m_Z/m_{A}\rightarrow 0$.
In particular, the couplings of the light Higgs boson $h$ are nearly equal to those of the SM Higgs boson. In a Supersymmetric framework, besides the higgs mediated LFV transitions, we have also LFV effects mediated by the gauginos through loops of neutralinos (charginos)- charged sleptons (sneutrinos).
On the other hand, the above contributions have different decoupling properties regulated by the mass of the heaviest scalar mass ($m_H$) or by the heaviest mass in the slepton gaugino loops ($m_{SUSY}$). However, in both cases, the effective operator for $l_i\rightarrow l_j\gamma$ is
$$
\frac{m_{l_i}}{m^2_{H,SUSY}}\,\,\bar{l_i}\sigma^{\mu\nu} l_j F_{\mu\nu}.
$$
In principle, the $m_{SUSY}$ and $m_H$ masses can be unrelated so,
we can always proceed by considering only the Higgs mediated effects
(assuming a relatively light $m_H$ and an heavy $m_{SUSY}$)
or only the gaugino mediated contributions (if $m_H$ is heavy).
So, taking into account the only Higgs-mediated effects,
we get the following branching ratio for $\mu\rightarrow e\gamma$
\footnote{
In a SUSY framework, the couplings between the scalar and the fermions 
are given by $-i(\sqrt 2 G_F)^{1/2}\tan\beta H\xi_{f} m_f\overline{f}f$ 
where the parameters $\xi_{f}$ are equal to one at tree level but
they can get large corrections from higher order effects.
For instance, $\xi_{b}$ gets contributions
from gluino-squark loops (proportional to $\alpha_{s}t_{\beta}$)
that enhance or suppress significantly the tree level value of $\xi_{b}$ \cite{hrs}.
In the $\xi_{\tau}$ case the leading one loop effects induced
by chargino-sneutrino contributions (proportional to $\alpha_{w}t_{\beta}$)
do not affect $\xi_{\tau}$ so significantly.
For simplicity's sake, we disregard the above factors in the following.}:
\beqa
\label{main}
Br(\mu\rightarrow e \gamma)
&\simeq&
\frac{3}{2}\frac{\alpha^3_{el}}{\pi^3}
\Delta_{21}^{2}t^{6}_{\beta}
\bigg[\sum_{f= b, \tau}\!\!N_f q^2_{f}
\frac{m^2_{f}}{M^2_{H}}\left(\log{\frac{m^2_{f}}{M^2_{H}}}+2\right)
\!-\!\frac{m^{2}_W}{M^{2}_{H}}\frac{F(a_{W})}{t_{\beta}}+
\nonumber\\\nonumber\\
&+&
\frac{N_c}{4}
\bigg(
q^{2}_{\tilde{t}}\,\frac{m_{t}\mu}{t_{\beta}M^2_{H}}\,\sin2\theta_{\tilde{t}}\,
h(x_{\tilde{t}H})-q^{2}_{\tilde{b}}\,\frac{m_{b}A_{b}}{M^2_{H}}\,
\sin2\theta_{\tilde{b}}\,h(x_{\tilde{b}H})
\bigg)
\bigg]^2
\nonumber\\\nonumber\\
&\simeq&
\frac{3}{2}\,\frac{\alpha^3_{el}}{\pi^3}\,
\Delta_{21}^{2}\,t^{4}_{\beta}\,
\bigg(\frac{m^{4}_W}{M^{4}_{H}}\bigg)\,\bigg(F(a_{W})\bigg)^{2}
\eeqa
where $a_W=m_{W}^{2}/m_H^2$, $x_{\tilde{f}H} = m_{\tilde{f}}^2/m_H^2 $, $\theta_{\tilde{t},\tilde{b}}$ are squarks mixing angles and the loop function $h(z)$ is given by:
\beq
\label{fmain}
h(z) = \int_0^1 dx \frac{x(1-x) \log{(z/x(1-x))}}{z-x(1-x)}.
\eeq
The asymptotic form of $h(z)$, which may be useful for an easy understanding
of the results, is given by:
\beq
 h(z) = \left\lbrace
\begin{array}{ll}
\label{f2main}
  -(\log z+2) \qquad \qquad & z \ll 1\\
  0.344 \qquad & z= 1\\
  \frac{1}{6 z}(\log z +\frac{5}{3}) \qquad & z \gg 1 .
\end{array}
\right.
\label{asymp}
\eeq
The first two terms of Eq.\ref{main} refer to two loop effects induced
by fermionic and $W$ loops, respectively, while the last term appears
only in a supersymmetric framework and it is relative to squark loops \cite{piledm}.
In the previous section, we have seen that $W$ effects dominate
over the fermionic ones. Moreover, being the $H$ and $A$ masses almost 
degenerate in the decoupling limit, the $H$ and $A$ contributions partially cancel themselves in the fermionic amplitude because of their opposite signs.
This is in contrast to the $W$ amplitude that turns out to be sensitive only
to $H$ effects.\\
As regards the squark loop effects, it is very easy to realize
that they are negligible compared to $W$ effects.
In fact, it is well known that Higgs mediated LFV can play a relevant
or even a dominant role compared to gaugino mediated LFV provided that
slepton masses are not below the $TeV$ scale while maintaining
the Higgs masses at the electroweak scale (and assuming large $t_{\beta}$ values).
In this context, it is natural to assume squark masses at least
of the same order as the slepton masses (at the $TeV$ scale)
so that $x_{\tilde{f}H}\gg1$. So, even for maximum squark mixings, namely for
$\sin2\theta_{\tilde{t},\tilde{b}}\simeq 1$,
and large $A_{b}$ and $\mu$ terms, two loop squark effects remain much below
the $W$ effects, as it is straightforward to check by Eqs.\ref{main},
\ref{fmain},\ref{f2main}.
In the decoupling limit, $\mu\rightarrow eee$ and $\mu Al\rightarrow e Al$
are still dominated by the two loop $\mu\rightarrow e\gamma^*$ amplitude
and by a three level Higgs exchange, respectively. 
Finally one gets the following relations:
\beqa
\frac{Br(\mu\rightarrow e \gamma)}{Br(\mu Al\rightarrow e Al)}\bigg|_{\rm{Higgs}} 
&\simeq& 10^{\,2}\,
\left(\frac{F(a_W)}{\tan\beta}\right)^{2}\,\,,\,\,\,\,
\frac{Br(\mu\rightarrow eee)}{Br(\mu \rightarrow e \gamma)}\bigg|_{\rm{Higgs}}
\simeq\alpha_{el}.
\eeqa
As we can note, the above predictions are exactly the same ones
we found in the non-decoupling limit and the numerical results are 
reported in fig.1. However, this property is no longer true for
processes associated to $e-\tau$ and $\mu-\tau$ transitions \cite{hmio}.
Let us now consider the one loop contributions to $\mu-e$
transitions arising from $(\Delta^{21})_{eff.}\!=\!\Delta^{23}\Delta^{31}$.
The corresponding $Br(\mu\rightarrow e\gamma)$ is:
\beqa
\label{2MIhiggs}
Br(\mu\rightarrow e \gamma)
&\simeq&
\frac{3}{2}\frac{\alpha_{el}}{\pi}
\bigg(\frac{m^2_{\tau}}{m^2_{A}}\bigg)^{2}t_{\beta}^{8}
\bigg[
\bigg(\frac{\delta m}{m_A}\log{\frac{m^2_{\tau}}{m^2_{A}}}\!+
\!\frac{1}{6}\bigg)\Delta_{L}^{23}\Delta_{L}^{31}+\nonumber\\
&+&
\bigg(\frac{m_{\tau}}{m_{\mu}}\bigg)
\bigg(\frac{\delta m}{m_A}\bigg)
\bigg(\log{\frac{m^2_{\tau}}{m^2_{A}}}\!+\!\frac{3}{2}\bigg)
\Delta_{R}^{23}\Delta_{L}^{31}
\bigg]^{2}+(L\!\leftrightarrow\!R)
\eeqa
where $\delta m\!=\!m_A-m_H$. The proportional term to $\Delta_{R}^{23}\Delta_{L}^{31}\sim\delta_{RR}^{23}\delta_{LL}^{31}$
in eq.16 is enhanced by an $m_{\tau}/m_{\mu}$ factor compared to the 
proportional term to $\Delta_{L}^{23}\Delta_{L}^{31}\sim\delta_{LL}^{23}\delta_{LL}^{31}$.
On the other hand, this enhancement is not effective in a Susy framework.
In fact, the upper bounds on $\delta_{RR}^{23}\delta_{LL}^{31}$
imposed by the gaugino mediated effects to $Br(\mu \rightarrow e \gamma)$
are stronger than those relative to $\delta_{LL}^{23}\delta_{LL}^{31}$
of the same $m_{\tau}/m_{\mu}$ factor \cite{1mio}, as we will discuss.
Differently from the non-decoupling limit case, now one loop effects with
two LFV couplings are suppressed by the mass splitting $\delta m/m_A$.
In a SUSY framework, if $\delta m/m_A\simeq 10\%$, $\Delta^{21}\sim 10^{-3}\delta^{21}$ and $\delta^{21}\sim \delta^{23}\delta^{31}$ we get
$Br^{\mu\rightarrow e\gamma}_{1-loop}$, roughly two-three order of magnitude below the $Br^{\mu\rightarrow e\gamma}_{2-loop}$ obtained from 
two loop effects with a single LFV coupling.
However, in a generic model II 2HDM, one loop effects can still provide the
major effects depending on the size of the $\Delta_{ij}$ terms.
\begin{figure}[ht]
\begin{tabular}{cc}
\includegraphics[scale=0.35]{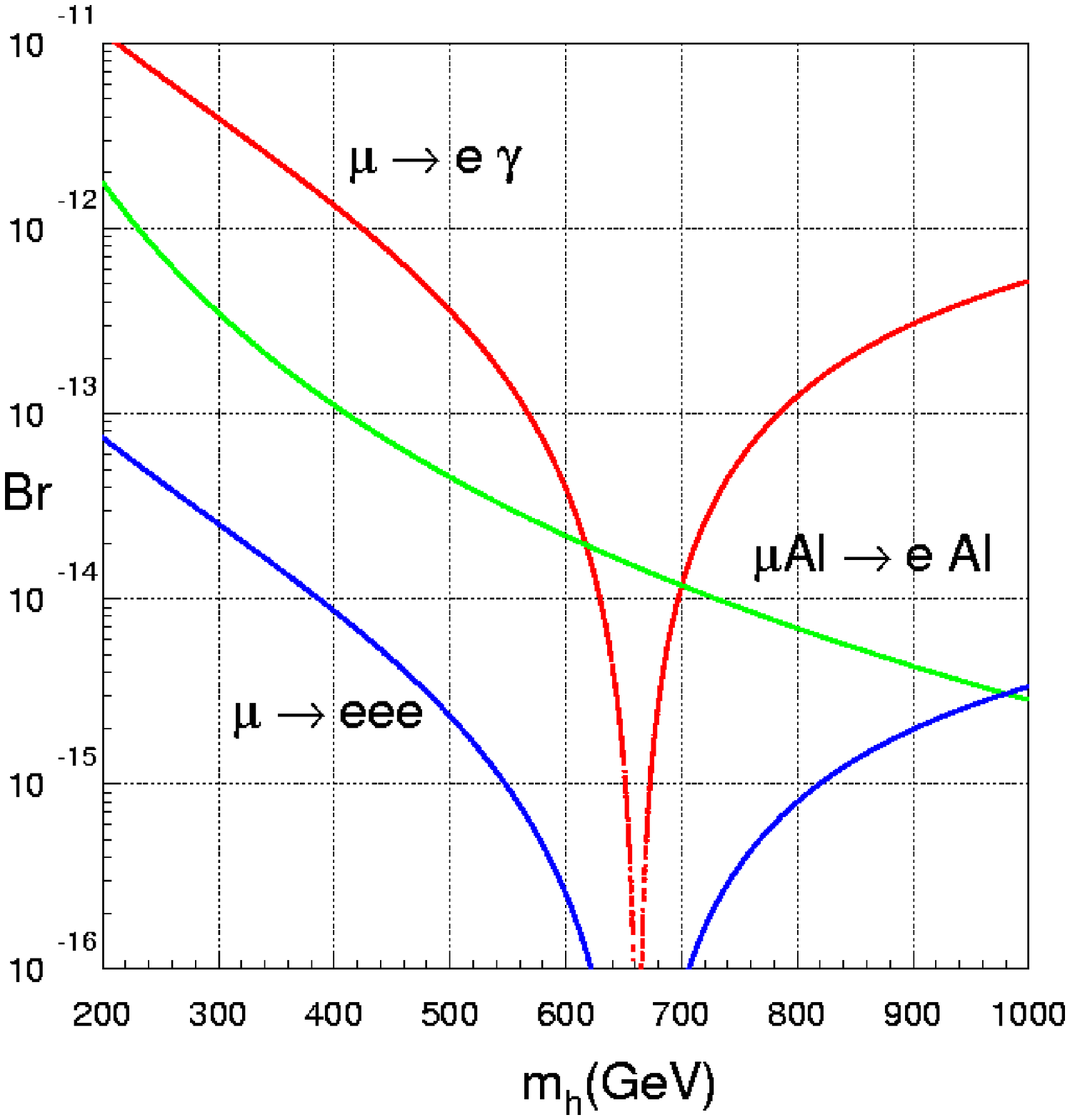} &
\includegraphics[scale=0.35]{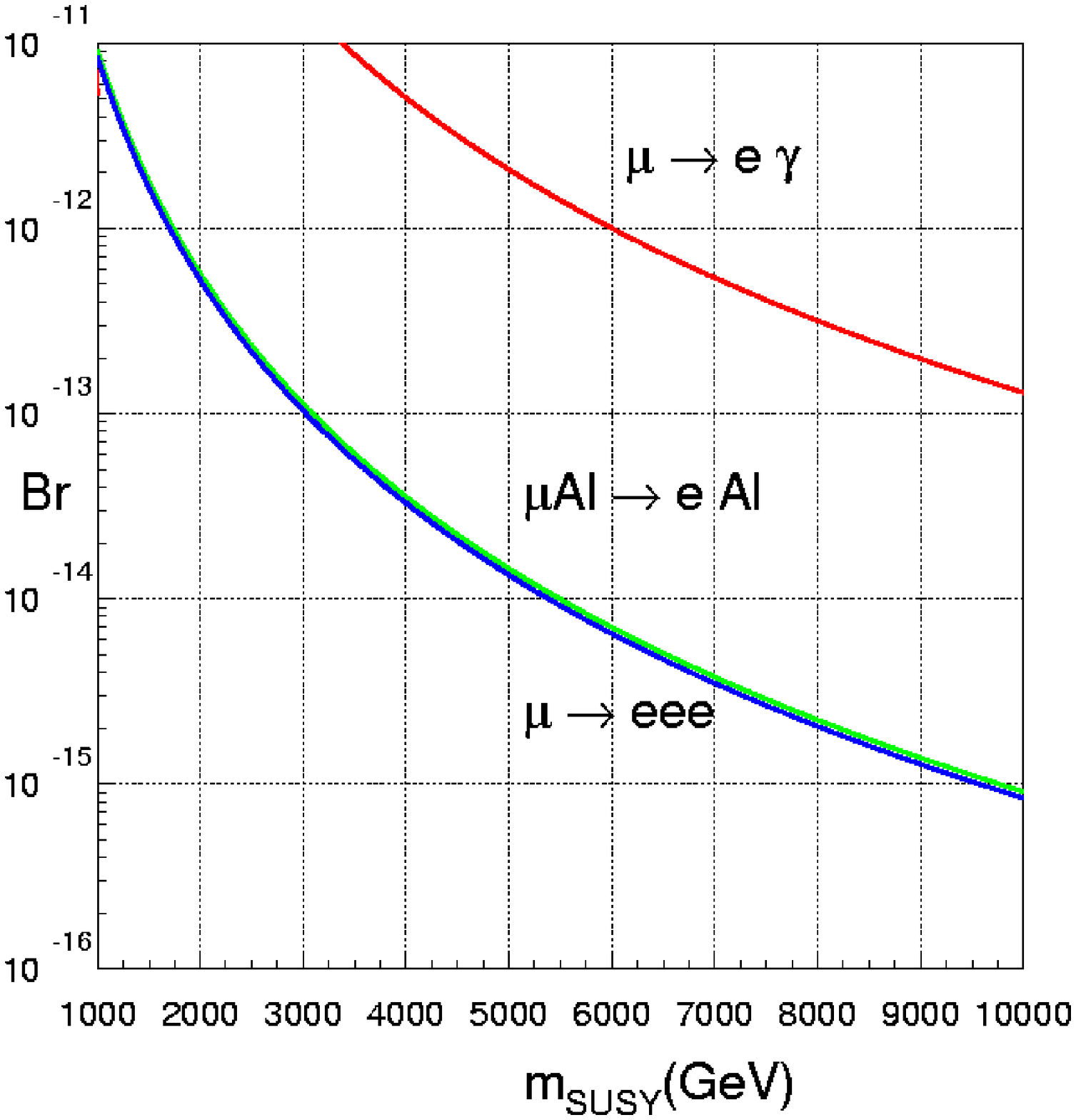}
\end{tabular}
\caption{Left: Branching ratios of $\mu\rightarrow e\gamma$, 
$\mu\rightarrow eee$ and $\mu Al\rightarrow e Al$ in the Higgs 
mediated LFV case vs the Higgs boson mass $m_h$.
In the decoupling (non-decoupling) limit $m_h$ refers to the heaviest
(lightest) Higgs boson mass.
Right: Branching ratios of $\mu\rightarrow e\gamma$, 
$\mu\rightarrow eee$ and $\mu Al\rightarrow e Al$ in the gaugino
mediated LFV case vs a common SUSY mass $m_{SUSY}$.
In both of figures we set $t_{\beta}=50$
and $\delta^{21}_{LL}=10^{-2}$ (that corresponds, in a generic 2HDM, 
to $\Delta_{L}^{21}\simeq 5\cdot 10^{-6}$).}
\end{figure}
%
In the following, we are interested to make a comparison between Higgs and
gaugino mediated LFV effects.
To this purpose let us report the branching ratio of $l_{i}\rightarrow l_{j}\gamma$ induced by the one loop exchange of neutralinos, 
charginos and sleptons
\beqa
\frac{BR(\mu\rightarrow e\gamma)}{BR(\mu\rightarrow e\nu_{\mu}\bar{\nu_e})} =
\frac{48\pi^{3}\alpha}{G_{F}^{2}}(|A_L|^2+|A_R|^2)\,,
 \nonumber
 \eeqa
where the $A_{L,R}$ amplitudes are given by
\beqa
\label{AL}
A_{L}=
\frac{\alpha_{2}}{4\pi}\delta_{LL}^{21} t_{\beta}
&\bigg[&
\mu M_{2}
\frac{\left(f_{2n}(a_{2L},b_L)
\!+\!f_{2c}(a_{2L},b_L)\right)}{m^4_{L}(M_{2}^2\!-\!\mu^2)}
\nonumber\\\nonumber\\
&+&
\tan^{\!2}\theta_{W} \mu M_{1}
\bigg(
\frac{-f_{2n}(a_{1L},b_L)}{m^4_{L}(M_{1}^2\!-\!\mu^2)}+
\frac{1}{(m^{2}_{R}-m^{2}_{L})}\cdot\nonumber\\\nonumber\\
&\bigg(&\frac{2f_{2n}(a_{1L})}{m^4_{L}}\!+\!\frac{1}{(m^{2}_{R}\!-\!m^{2}_{L})}
\bigg(\frac{f_{3n}(a_{1R})}{m^2_{R}}\!-\!\frac{f_{3n}(a_{1L})}{m^2_{L}}\bigg)
\!\bigg)
\!\bigg)
\!\bigg],
\eeqa
\beqa
\label{AR}
A_R=\frac{\alpha_{1}}{4\pi}\delta_{RR}^{21} t_{\beta}\mu M_{1}
&\bigg[&
\frac{2f_{2n}(a_{1R},b_R)}{m^4_{R}(M_{1}^2\!-\!\mu^2)}\!+\!
\frac{1}{(m^{2}_{L}\!-\!m^{2}_{R})}\cdot\nonumber\\\nonumber\\
&\bigg(&\frac{2f_{2n}(a_{1R})}{m^4_{R}}+\frac{1}{(m^{2}_{L}\!-\!m^{2}_{R})}
\bigg(\frac{f_{3n}(a_{1L})}{m^2_{L}}\!-\!
\frac{f_{3n}(a_{1R})}{m^2_{R}}\bigg)\!\bigg)
\!\bigg]\,,
\eeqa
respectively, and $a_{1L,2L}=M^{2}_{1,2}/m^{2}_{L}$, 
$a_{1R}=M^{2}_{1}/m^{2}_{R}$ and $b_{L,R}=\mu^2/m^{2}_{L,R}$.
The loop functions $f_{i(c,n)}(x)$'s are such that
$f_{i(c,n)}(x,y)=f_{i(c,n)}(x)-f_{i(c,n)}(y)$ with.\\
$$
f_{2n}(a)=\frac{-5a^2+4a+1+2a(a+2)\ln a}{4(1-a)^4}\qquad
f_{3n}(a)=\frac{1+2a\ln a-a^2}{2( 1-a)^3}
$$
$$
f_{2c}(a)=\frac{-a^2-4a+5+2(2a+1)\ln a}{2(1-a)^4}\,.
$$
As we can see from Eq.~(\ref{AL}), the $A_L$ amplitude
includes both $U(1)$ (the terms proportional to
$\tan^{\!2}\theta_{W}$) and $SU(2)$ type contributions.
The $U(1)$ contributions correspond to pure $\tilde B$ exchange,
with chirality-flip in the internal fermion line or to 
$\tilde B-\tilde H^0$ exchange with chirality flip realized
at the Yukawa vertex.
For the $SU(2)$ case, we have $\tilde W-\tilde H$ exchange both 
for charginos and for neutralinos.
However, given that $\tilde W$ fields do not couple to right-handed fields,
pure $\tilde W$ exchange can not mediate any contribution with internal
sfermion line chirality flip in contrast to the $U(1)$ case.
On the contrary, the $A_R$ amplitude receives only $U(1)$
contributions.
As regard the $A_R$ amplitude, we observe that it suffers from some 
cancellations among different contributions in regions of the parameter space.
The origin of these cancellations is the destructive interference between 
the contributions coming from the $\tilde B$ (with internal chirality flip) 
and $\tilde B \tilde H^{0}$ exchange.
It is easy to check numerically that these contributions,
have opposite sign in all the parameter space.
On the other hand the same type of contributions in the $\delta_{LL}$ 
case have the same sign. The reason for this difference is the opposite sign 
in the hypercharge of $SU(2)$ doublets and singlets.

Finally, we observe that, if the SUSY model contains both $\delta_{LL}^{23}$ 
and $\delta_{RR}^{31}$ MI types, we get an additional contribution 
so that $A^{tot}_{L}=A_{L}+A^{'}_{L}$ with $A^{'}_{L}$ given by:
\beqa
\label{2MIgauge}
A^{'}_{L}=
&-&\frac{\alpha_{1}}{2\pi}\,
\left(\frac{m_{\tau}}{m_{\mu}}\right)\,\mu M_{1}t_{\beta}
\frac{\delta_{LL}^{23}\delta_{RR}^{31}}{(m^{2}_{L}\!-\!m^{2}_{R})^2}\cdot
\nonumber\\\nonumber\\
&\cdot&
\bigg[
\frac{f_{2n}(a_L)}{m^4_{L}}+\frac{f_{2n}(a_R)}{m^4_{R}}+
\frac{1}{(m^{2}_{R}\!-\!m^{2}_{L})}
\bigg(\frac{f_{3n}(a_R)}{m^2_{R}}\!-\!\frac{f_{3n}(a_L)}{m^2_{L}}\bigg)
\bigg].
\eeqa
A particularly interesting feature of the above amplitude is the $m_{\tau}/m_{\mu}$ enhancement with respect to the usual Bino-like mediated processes. 
This is due to the implementation of the chirality flip in the internal sfermion line through $\delta^{LR}_{33}\sim m_{\tau}\mu\tan\beta$ and not by 
$\delta^{LR}_{22}\sim m_{\mu}\mu\tan\beta$, as usual.
The $A^{'}_{R}$ amplitude, relative to $\delta^{RR}_{23}\delta^{LL}_{31}$,
is simply obtained by $A^{'}_{R}=A^{'}_{L}(L\leftrightarrow R)$.
The contribution reported in Eq.\ref{2MIgauge} has to be compared 
with the second term of Eq.\ref{2MIhiggs}
that is the analog contribution in the Higgs mediated LFV case.

We stress that, in Eq.\ref{2MIgauge}, we have not included contributions
proportional to $\delta^{RR,LL}_{23}\delta^{RR,LL}_{31}$ because they
are generally suppressed (or at most comparable) compared to those proportional 
to $\delta^{RR,LL}_{21}$.
On the contrary, in Eq.\ref{2MIhiggs}, terms proportional to $\Delta^{RR,LL}_{23}\Delta^{RR,LL}_{31}$
were retained because enhanced by a $(m_{\tau}/m_{\mu})^2$ factor compared 
to the corresponding effects proportional to $\Delta^{RR,LL}_{21}$.

The processes $\mu\rightarrow eee$ and $\mu$--$e$
conversion in Nuclei get contributions not only from penguin-type diagrams
(both with photon or Z-boson exchange) but also from box-type diagrams.
In fact, the dipole $\mu\rightarrow e \gamma^{*}$ contribution in these
processes is also given by Eqs.~\ref{AL},\ref{AR} and
therefore is enhanced by a $\tan\beta$ factor. On the other hand the other
contributions, monopole or boxes, are not proportional to $\tan\beta$.
So the dipole contributions usually dominate specially in the large
$\tan\beta$ regime and one can find the simple theoretical relations
\beqa
\label{relations}
\frac{Br(\mu\!\rightarrow\!eee)}{Br(\mu\!\rightarrow\!e\gamma)}
\bigg|_{\rm{Gauge}}
\simeq\!
\frac{Br(\mu - e {\rm\ in \ Ti})}{Br(\mu\!\rightarrow\!e\gamma)}
\bigg|_{\rm{Gauge}}
\simeq\! 
\alpha_{el}
\eeqa
In order to make the comparison between Higgs and Chargino mediated LFV
effects as simple as possible, let us consider the simple case where
all the susy particles are degenerate. In this case, it turns out that
$$
\Delta^{21}_{L}\sim \frac{\alpha_{2}}{24\pi}\delta^{21}_{LL}\,,
$$
\beqa
BR(\mu\rightarrow e\gamma)\bigg|_{\rm{Gauge}} =
\frac{2\alpha_{el}}{75\pi}\bigg(1\!+\!\frac{5}{4}\tan^{2}\theta_{W}\bigg)^2
\bigg(\frac{m^{4}_{W}}{m^{4}_{SUSY}}\bigg)
\bigg(\delta^{21}_{LL}\bigg)^2 t^{2}_{\beta}\,,
\nonumber
 \eeqa
\beq
Br(\mu\rightarrow e \gamma)\bigg|_{\rm{Higgs}}
\simeq
\frac{3}{2}\,\frac{\alpha^3_{el}}{\pi^3}\,
\bigg(\frac{\alpha_{2}}{24\pi}\bigg)^{2}
\bigg(\frac{m^{4}_W}{M^{4}_{H}}\bigg)\,
\bigg(F(a_{W})\bigg)^{2}
\bigg(\delta^{21}_{LL}\bigg)^2\,t^{4}_{\beta}\,.
\eeq
In Fig.1 we report the branching ratios of the examined processes 
as a function of the heaviest Higgs boson mass $m_H$
(in the Higgs LFV mediated case)
or of the common susy mass $m_{SUSY}$ (in the gaugino LFV mediated case).
We set $t_{\beta}=50$ and we consider the PMNS scenario as discussed
in Section 2 so that $(\delta^{21}_{LL})_{PMNS}\simeq 10^{-2}$.
Subleading contributions proportional to 
$(\delta^{23}_{LL(RR)}\delta^{31}_{RR(LL)})_{PMNS}$
(see Eqs.\ref{2MIhiggs},\ref{2MIgauge}) were neglected since, 
in the PMNS scenario, it turns out that  
$(\delta^{23}_{LL(RR)}\delta^{31}_{RR(LL)})_{PMNS}/(\delta^{21}_{LL})_{PMNS}\simeq 10^{-3}$
\cite{masivemp}.
As we can see from Fig.1, Higgs mediated effects start being competitive
with the gaugino mediated ones when $m_{SUSY}$ is roughly one order of
magnitude larger then the Higgs mass $m_H$.
Moreover, we stress that, both in the gaugino and in the Higgs mediated cases,
$\mu\rightarrow e\gamma$ gets the largest effects.
In particular, within the PMNS scenario,
it turns out that Higgs mediated $Br(\mu\rightarrow e\gamma)\sim 10^{-11}$ 
when $m_{H}\sim 200\rm{GeV}$ and $t_{\beta}=50$,
that is just closed to the present experimental resolution.

\section{Conclusions}

In this letter we have studied Higgs-mediated LFV $e-\mu$ transitions 
in 2HDM and Supersymmetry frameworks.
The sources of LFV were parametrized in a model independent way in order 
to be as general as possible.
In particular, we have analyzed $\mu\rightarrow e\gamma$,
$\mu\rightarrow eee$ and $e-\mu$ conversion in nuclei finding that
$\mu\rightarrow e\gamma$ is generally the most
sensitive channel to probe Higgs-mediated LFV.
Analytical expressions for the rates of the above processes and their correlations 
have been established up to two loop level.
Particular emphasis was given to the correlations among the processes
as an important signature of the theory.
In fact, while it is rather difficult to predict
the absolute branching ratio value for any given process
(depending on the amount of LFV sources and on the mass spectrum),
possible correlations with other processes seem to be a more powerful tool 
to disentangle different scenarios.
In this respect, experimental improvements in all the examined
$e-\mu$ transitions would be very welcome.
On the other hand, we have shown that the Higgs-mediated contributions to
LFV processes can be within the present or upcoming experimental
resolutions and provide an important chance to detect new physics
beyond the Standard Model.\\

\textbf{Acknowledgments:}
I thank A.Brignole, G.Isidori, R.Kitano and A.Masiero for useful discussions.

\end{document}